 \begin{filecontents*}{example.eps}
\RequirePackage{fix-cm}
\documentclass{svjour3}                     
\smartqed  
\usepackage{graphicx}
\usepackage[english]{babel}
\usepackage{amsmath}
\usepackage{mathrsfs}
\usepackage{amssymb}

\usepackage{amsthm}
\usepackage{enumitem}
\usepackage{bm}
\usepackage{float}
\usepackage{ wasysym }
\usepackage{color, colortbl}
\definecolor{Gray}{gray}{0.9}
\usepackage{verbatim}
\usepackage{polynom}
\usepackage{multicol}

\usepackage{subfig}
\usepackage{epstopdf}

\usepackage[square, sort, numbers]{natbib}
\usepackage{multirow}

\newcommand{\bLozenge}{\mathbin{\blacklozenge}}
 \usepackage[table,xcdraw]{xcolor}
 \usepackage{hyperref}
\begin{document}

\title{Modeling in higher dimensions to improve diagnostic testing accuracy: theory and examples for multiplex saliva-based SARS-CoV-2 antibody assays
}
\subtitle{}

\titlerunning{Modeling in higher dimensions}        

\author{Rayanne A. Luke$^{1,2}$ \and Anthony J. Kearsley$^2$ \and Nora Pisanic$^3$ \and Yukari C. Manabe$^4$ \and David L. Thomas$^4$ \and Christopher D. Heaney$^{3,5,6}$ \and Paul N. Patrone$^2$
}


\institute{
             $^1$Johns Hopkins University, Whiting School of Engineering, Department of Applied Mathematics and Statistics, Baltimore, MD  
             \and             
$^2$National Institute of Standards and Technology, Applied and Computational Mathematics Division, Gaithersburg, MD 
\and
$^3$Johns Hopkins University, Bloomberg School of Public Health, Department of Environmental Health and Engineering, Baltimore, MD 
\and
$^4$ Johns Hopkins University, School of Medicine, Department of Medicine, Baltimore, MD 
\and
$^5$ Johns Hopkins University, Bloomberg School of Public Health, Department of Internal Health, Baltimore, MD 
\and
$^6$ Johns Hopkins University, Bloomberg School of Public Health, Department of Epidemiology, Baltimore, MD \\
\textbf{Corresponding author}: Rayanne A. Luke, rluke@jhu.edu \\
Senior authors: Paul N. Patrone and Christopher D. Heaney. Questions about assay design can be directed to Christopher D. Heaney at cheaney1@jhu.edu.
}

\date{2022}

\maketitle

\begin{abstract}

The severe acute respiratory syndrome coronavirus 2 (SARS-CoV-2) pandemic has emphasized the importance and challenges of correctly interpreting antibody test results. Identification of positive and negative samples requires a classification strategy with low error rates, which is hard to achieve when the corresponding measurement values overlap. Additional uncertainty arises when classification schemes fail to account for complicated structure in data. We address these problems through a mathematical framework that combines high dimensional data modeling and optimal decision theory. Specifically, we show that appropriately increasing the dimension of data better separates positive and negative populations and reveals nuanced structure that can be described in terms of mathematical models. We combine these models with optimal decision theory to yield a classification scheme that better separates positive and negative samples relative to traditional methods such as confidence intervals (CIs) and receiver operating characteristics. We validate the usefulness of this approach in the context of a multiplex salivary SARS-CoV-2 immunoglobulin G assay dataset. 
This example illustrates how our analysis: (i) improves the assay accuracy (e.g. lowers classification errors by up to 42 \% compared to CI methods); (ii) reduces the number of indeterminate samples when an inconclusive class is permissible (e.g. by 40 \% compared to the original analysis of the example multiplex dataset); and (iii) decreases the number of antigens needed to classify samples. 
Our work showcases the power of mathematical modeling in diagnostic classification and highlights a method that can be adopted broadly in public health and clinical settings.

\end{abstract}

\section{Introduction}

Antibody testing has become a crucial public health tool during the severe acute respiratory syndrome coronavirus 2 (SARS-CoV-2) pandemic. Interpretation of serology results typically uses mathematical analyses that classify samples as positive, negative, or indeterminate. Conventional classification strategies include confidence interval (CI)-based schemes or receiver operating characteristics (ROC) \cite{jacobson1998validation,florkowski2008sensitivity}. The accuracies of both methods suffer from assumptions that fail to account for the structure of data.

Despite having a fundamental impact on classification, this concept of data structure is rarely acknowledged and requires further consideration. For example, CI and ROC-based methods do not address questions such as: how are positive and negative populations clustered? In what direction does each population’s data spread out? What shapes outline the measurement values of each population? Throughout we refer to such ideas as the \textit{structure} of data. 

Equally important, neither CIs nor ROC thoroughly characterize the structure of positive and negative populations relative to each other. CIs for negative populations are decoupled from information about positive samples. Specifically, they label measurements outside a fixed number of standard deviations (3$\sigma$) from the negative sample mean as positive \cite{algaissi2020sars,grzelak2020comparison,hachim2020orf8}. However, low probability of being negative does not imply high probability of being positive. Moreover, this choice implicitly assumes that 99.7 \% of negatives fall within the CI, or that a Gaussian model fits the data, which may be unreasonable. ROC is a graphical method with a straightforward interpretation as minimizing error, but it fails to account for how the degree of overlap between positive and negative measurements changes with prevalence. These observations suggest that more accurate classification methods can be realized by explicitly quantifying the structure of data.

One strategy to overcome certain limitations of CIs and ROC is to build mathematical modeling-based classification schemes, which can leverage structural properties of data to improve accuracy. Probability models can be formulated to quantify phenomena, such as: (i) the degree to which positive samples have higher antibody levels than negatives; (ii) statistical correlation of data; and (iii) the outline of data described in terms of shapes like spheres or cones. Several previous works have used mathematical modeling in diagnostic classification \cite{patrone2021classification,bottcher2022statistical}. A recent approach applied a combination of modeling and optimal decision theory to antibody testing and proved optimality for binary classification \cite{patrone2021classification}; \cite[see][Chapter 3]{williams2006gaussian}. Another study built a statistical model for binary classification that accounted for sample bias and used either antibody or viral-load tests \cite{bottcher2022statistical}. 

In contrast to previous work, we use high-dimensional mathematical modeling to address the shortcomings of CIs and ROC. We first observe that appropriately increasing the dimension of data—i.e., measuring IgG binding to relevant additional antigens, or, in a multiplex assay, taking advantage of previously unused information—can reveal nuanced structure and better separate positive and negative populations. Next, we construct conditional probability models for the corresponding measurements. We use an optimal classification scheme \cite{patrone2021classification} to improve accuracy and decrease the number of indeterminate test results when such a class is permissible. We validate these ideas by constructing three-dimensional (3D) models for a SARS-CoV-2 immunoglobulin G (IgG) assay \cite{pisanic2020covid,randad2021durability}.

By construction, our 3D modeling classification is more accurate than CI methods, because we both adapt to the structure of the data and use an optimal method. ROC does not have an analog to our results because the technique is unusable beyond one dimension \cite[see][]{patrone2021classification}. Further, antibody assays can have a narrow linear detection range that interferes with the interpretation of statistical CIs; in contrast, our method makes no assumption about linearity of the detector, and we can account for such (and related) effects as needed.

A key result of our work is that increased data separation in higher dimensions decreases the number of indeterminate test results while improving classification accuracy. As an unexpected bonus, our higher dimensional work allows us to improve upon the performance in a related work \cite{patrone2022holdout} while reducing the number of measurement targets in the analysis from eight to three. We attribute this improvement to the increased data separation of working in 3D and the fidelity of our models to the structure revealed in higher dimensions. 

A key goal of this manuscript is to be accessible to the broader clinical community. As such, we provide intuitive examples and highlight core ideas, leaving technical details and most equations for the Appendix. In the Materials and Methods section, we introduce the multiplex oral fluid (hereafter, salivary) SARS-CoV-2 IgG assay data and provide a straightforward understanding of our 3D models with figures. We display the optimal classification domains and compare our results to CI methods and the original analysis \cite{pisanic2020covid} in the Results section; this is followed by further analysis, limitations, and extensions in the Discussion.

\section{Materials and Methods}

\subsection{Data Introduction}

We consider a multiplex salivary SARS-CoV-2 IgG assay that measures anti-IgG specific to three domains of SARS-CoV-2: nucleocapsid (N) protein, receptor binding domain (RBD), and full spike (S) protein \cite{pisanic2020covid,randad2021durability}. Measurements are reported as median fluorescence intensities (MFIs). The dataset is separated into training and test populations following the same delineation as the original analysis \cite{pisanic2020covid,randad2021durability} for which the true classes are treated as known and unknown, respectively. Notably, the original assay used seven SARS-CoV-2 antigen-specific IgG targets; we will show that our modeling achieves better accuracy using only one N and one RBD antigen. We empirically select these antigens as the pair with the greatest separation as defined by the silhouette coefficient \cite{kogan2007introduction}. In this context, the silhouette coefficient measures how close a negative sample is to the mean of the negative population relative to how close it is to the mean of the positive population. (Questions of optimal down-selection are reserved for future work.)

Positive samples were confirmed via quantitative polymerase chain reaction tests and were collected more than fourteen days post-coronavirus disease of 2019 (COVID-19) symptom onset. Negative samples were collected prior to the pandemic and have small signals, potentially arising from cross-reactivity with other coronaviruses such as 229E, NL63, OC43, and HKU1. Data in the original analysis \cite{pisanic2020covid,randad2021durability} were labeled indeterminate if samples classified as SARS-CoV-2 IgG negative contained salivary total IgG less than or equal to 15 $\mu$g/mL as measured by an enzyme-linked immunosorbent assay (ELISA). In this way, salivary total IgG was used as a measure of sample adequacy. Additionally, some samples were labeled indeterminate in the original analysis due to low sample volume and instrument error. In contrast, we establish our indeterminate class by a variation of a method that holds out samples with the lowest probability of being correctly classified \cite{patrone2022holdout}. This is discussed in the Indeterminate Class subsection. 

The two antibody targets are represented as a measurement double $\bm{\hat{r}}=(\hat{x} ,\hat{y } )$. The measurements are transformed to a logarithmic scale, via the equation
\begin{equation}
x = \log_2(\hat{x} + 2) - 1,
\label{eq:1}
\end{equation}
with a similar transformation defining the variable $y$. This transformation is a modeling choice that puts the data on the scale of bits and separates the data well. We denote the corresponding variables by $\bm{r}=(x,y)$. Note that $\bm{r}$ can be interpreted as an ordered pair in 2D. We plot N vs. RBD in Figure \ref{fig:1}a. The positive population is indicated by red \textbf{X}s and the negatives by blue $\bLozenge$s. The green box indicates the 3$\sigma$ CIs for the negative population, which do not model the correlation structure of the data. Figure \ref{fig:1}a shows that in two dimensions some positive and negative samples overlap. 

\begin{figure}
\centering
\includegraphics[scale=.45]{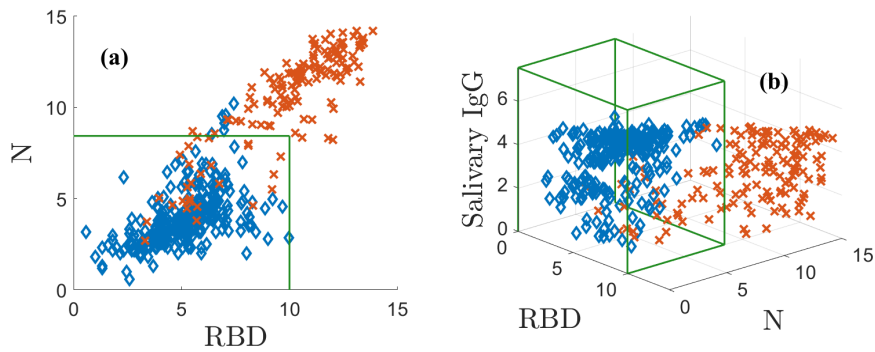}
\caption{Positive (red \textbf{X}) and negative (blue $\bLozenge$s) training antibody data. (a) N plotted against RBD, (b) the ELISA-based total IgG is added as the vertical axis. The green boxes in (a) and (b) are the negative sample mean plus 3$\sigma$ confidence intervals. \href{https://livejohnshopkins-my.sharepoint.com/:f:/g/personal/rluke3_jh_edu/Emfsn6KpxtBLv9rg0VmTkFEBWn6PspYkfYTmaYyvuEJD3Q?e=eFNhOA}{Supplemental Figure 1} shows an animation of (b).}
\label{fig:1}
\end{figure}

In Figure \ref{fig:1}b, adding the ELISA-based salivary total IgG values as a third dimension separates the data by lifting it into 3D. This additional measurement dimension must be related to the assay in a meaningful way to be useful. The three antibody measurements are transformed as above and represented as a measurement triple $\bm{r}=(x,y,z)$, which can be thought of as a point in 3D space. In doing so, many positively and negatively classified samples are pulled away from each other, significantly reducing overlap. The N and RBD levels of the negative data are lower than the positive values. The positive data are distributed roughly along a diagonal from the origin to the upper right corner, where all three of N, RBD, and the ELISA-based salivary total IgG are large. By eye, the different classes are much better separated in 3D than 2D, although there remains some overlap. Even in 3D, the CIs do not capture the structure of the data. The next section shows how modeling overcomes this problem. 

\subsection{Probability Models}

Our models predict the probability that a \textit{known} positive or \textit{known} negative sample yields a triple $\bm{r}$ of RBD, N, and ELISA-based salivary total IgG measurements. The training data is plotted with the models in Figure \ref{fig:2}. Regions of constant color are equal probability contours; the color corresponds to the positive (yellow) and negative (purple) models. The inner, darker volumes are regions of high probability that a given population yields a specific measurement value. Figure \ref{fig:2} shows the probability model contours for positive and negative populations along with the training data. The models quantify the structure in the two populations, which will later allow us to better classify samples. Figure \ref{fig:2} and its corresponding animation (see \href{https://livejohnshopkins-my.sharepoint.com/:f:/g/personal/rluke3_jh_edu/Emfsn6KpxtBLv9rg0VmTkFEBWn6PspYkfYTmaYyvuEJD3Q?e=eFNhOA}{Supplemental Figure 2}) are powerful illustrations of the model fits to the data and a useful way to understand its structure. We denote the positive model by $P(\bm{r})$ and the negative model by $N(\bm{r})$. Detailed mathematical descriptions of models are provided in Appendix \ref{sec:app_a}. 

\begin{figure}
\centering
\includegraphics[scale=.59]{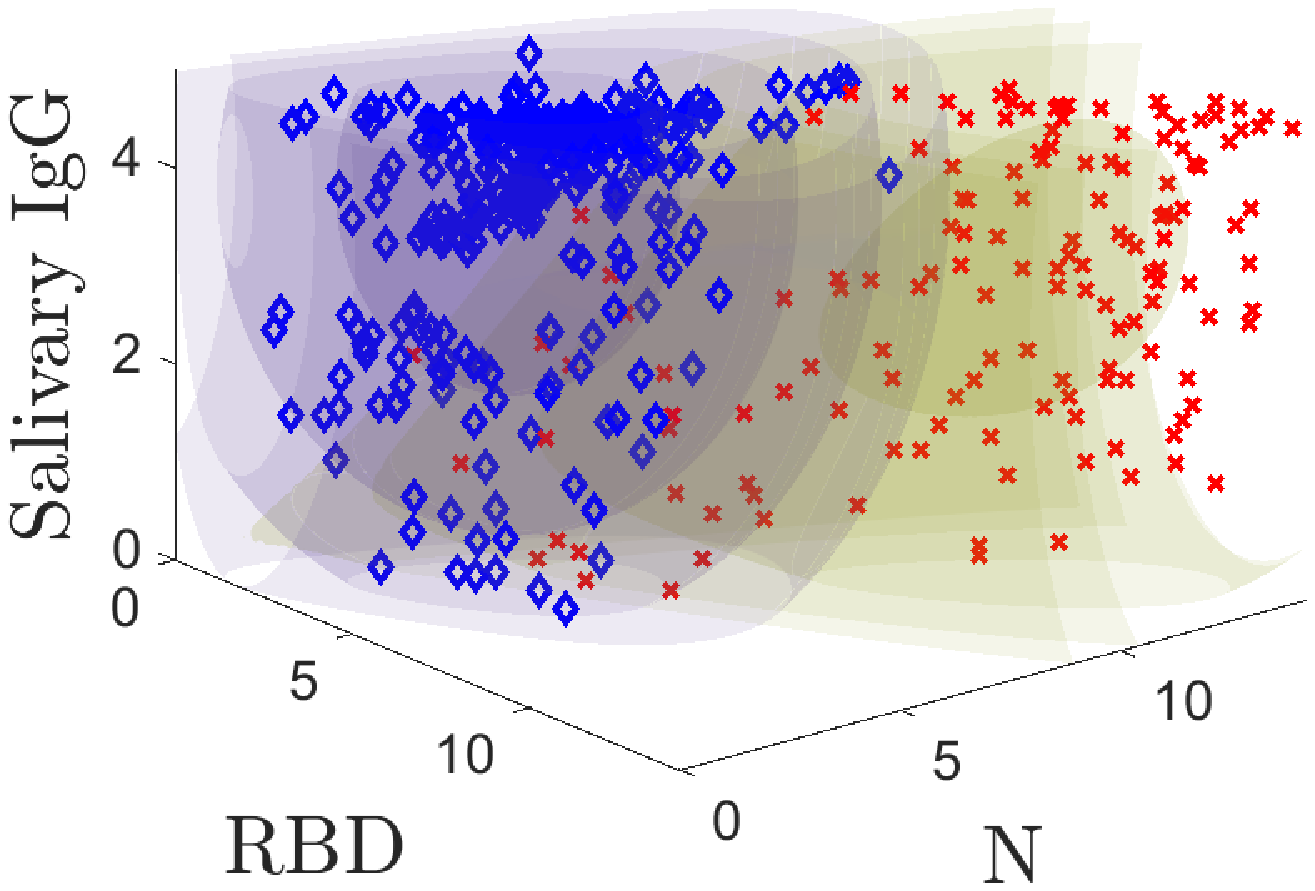}
\caption{3D probability model plotted along with the training data. Positive samples are indicated by red \textbf{X}s and negatives with blue $\bLozenge$s. \href{https://livejohnshopkins-my.sharepoint.com/:f:/g/personal/rluke3_jh_edu/Emfsn6KpxtBLv9rg0VmTkFEBWn6PspYkfYTmaYyvuEJD3Q?e=eFNhOA}{Supplemental Figure 2} shows an animation.}
\label{fig:2}
\end{figure}

\subsection{Classification}

Denote the prevalence of previously-infected individuals in the population by $q$; the fraction of uninfected individuals is $1-q$. The probability that a random sample is both positive and has measurement $\bm{r}$ is $qP(\bm{r})$; the probability that a random sample is both negative and has measurement $\bm{r}$ is $(1-q)N(\bm{r})$. A measurement equally likely to be positive or negative satisfies the equation:
\begin{equation}
qP(\bm{r}) = (1 - q) N(\bm{r}).
\label{eq:2}
\end{equation}
\textit{This equation defines a boundary in 3D that is analogous to cutoff values routinely used for single-antigen assays.} If the probability of being positive is greater, the measurement is classified as positive. If the probability of being negative is greater, the measurement is classified as negative. This classification scheme maximizes the accuracy, which we define as the prevalence-weighted combination of sensitivity and specificity; see \cite{patrone2021classification} for the objective function (Eq. 5) and justification of these statements.

\subsection{Indeterminate Class}

	We construct our indeterminate class by holding out samples with the lowest probability of being correctly classified. This concept of \textit{local accuracy} \cite{patrone2022holdout} gives the probability conditioned on a measurement value that the corresponding sample is correctly classified by optimal classification domains. Equations are given in Appendix \ref{sec:app_c}. The method \cite{patrone2022holdout} solved an optimization problem to determine the minimal number of samples to hold out to achieve a desired classification accuracy. We take a simpler approach by identifying a local accuracy threshold up to which we hold out samples so that the empirical classification accuracy reaches a desirable level. In higher dimensions, the modeling allows us to hold out fewer samples than both the original analysis \cite{pisanic2020covid} and the local accuracy work \cite{patrone2022holdout} while improving classification accuracies. This capitalizes on the data separation abilities of additional measurement dimensions. In the Results section, we describe our optimal classification domains for our example dataset and compare our classification sensitivities, specificities, and accuracies to those of CI methods and the two aforementioned papers.
	
	\section{Results}
	
	The models built on the training data are used to create the optimal classification domains; the classification boundary is given by Eq. \ref{eq:2}. Figure \ref{fig:3} shows the optimal positive (yellow) and negative (purple) classification domains for the SARS-CoV-2 IgG data. See \href{https://livejohnshopkins-my.sharepoint.com/:f:/g/personal/rluke3_jh_edu/Emfsn6KpxtBLv9rg0VmTkFEBWn6PspYkfYTmaYyvuEJD3Q?e=eFNhOA}{Supplemental Figures 3 and 4} for animations of Figure \ref{fig:3}.\textit{ The curved boundary between the negative and positive optimal domains reflects the structure and separation of the sample classes.}
	
	\begin{figure}
	\centering
	\includegraphics[scale=.45]{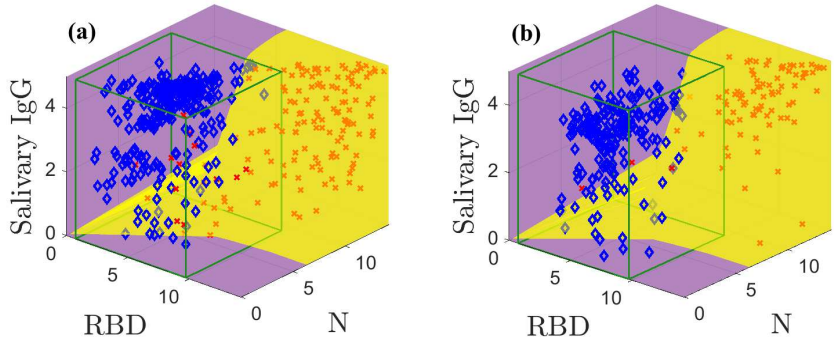}
	\caption{Optimal classification domains for the training (a) and test (b) data. Positive samples are indicated by red \textbf{X}s and negatives with blue $\bLozenge$s. The green boxes drawn in (a) and (b) are the negative sample mean plus 3$\sigma$ confidence intervals. \href{https://livejohnshopkins-my.sharepoint.com/:f:/g/personal/rluke3_jh_edu/Emfsn6KpxtBLv9rg0VmTkFEBWn6PspYkfYTmaYyvuEJD3Q?e=eFNhOA}{Supplemental Figures 3 and 4} show animations.}	
	\label{fig:3}
	\end{figure}

Our classification accuracy directly depends on this fidelity of our models to the data. Figure \ref{fig:3}a shows results for the training data with known prevalence $q$. Our model-based classification distinguishes between high and low salivary total IgG concentration for the same anti-N and anti-RBD IgG level. For example, in the middle of the plot we correctly classify negative samples with high total IgG concentrations and positive samples with low total IgG concentrations. Further, the optimal positive domain captures many positive samples even though it is nearly impossible to separate some negatives and positives that cluster along the diagonal. In contrast, for this dataset a CI categorizes samples with a fixed same N and RBD measurements and a variable ELISA-based salivary total IgG as members of the same class, leading to higher error rates. Our method admits four false positives and 16 false negatives; the CI yields six false positives and 26 false negatives. Despite inaccuracies, the modeling reduces false classifications by 37.5 \%.  

Figure \ref{fig:3}b shows the classification scheme applied to the test data using an estimated prevalence $\hat{q}$. This estimated prevalence can be computed using the models \textit{without classification} (see Appendix \ref{sec:app_b} and \cite{patrone2021classification}). Our method yields seven false classifications; the CI admits ten. The optimal classification domains are similar to those in Figure \ref{fig:3}a due to slight randomness in the prevalence estimate. 

Figure \ref{fig:4} shows the optimal classification domains for the models with indeterminate samples excluded in the white domain; \href{https://livejohnshopkins-my.sharepoint.com/:f:/g/personal/rluke3_jh_edu/Emfsn6KpxtBLv9rg0VmTkFEBWn6PspYkfYTmaYyvuEJD3Q?e=eFNhOA}{Supplemental Figures 5 and 6} are animations of Figure \ref{fig:4}. The indeterminate domain removes many samples in regions of high overlap between populations. Specifically, many positive samples with low N and RBD values are held out as these previously overlapped with many negative samples. The indeterminate domain describes the characteristic minimum separation between positive and negative samples that is needed to classify them with confidence; holding out samples that cannot be classified with sufficient accuracy increases sensitivity and specificity of the remaining data. The CI-based classification is also applied to the samples without indeterminates for comparison.

\begin{figure}
\centering
\includegraphics[scale=.46]{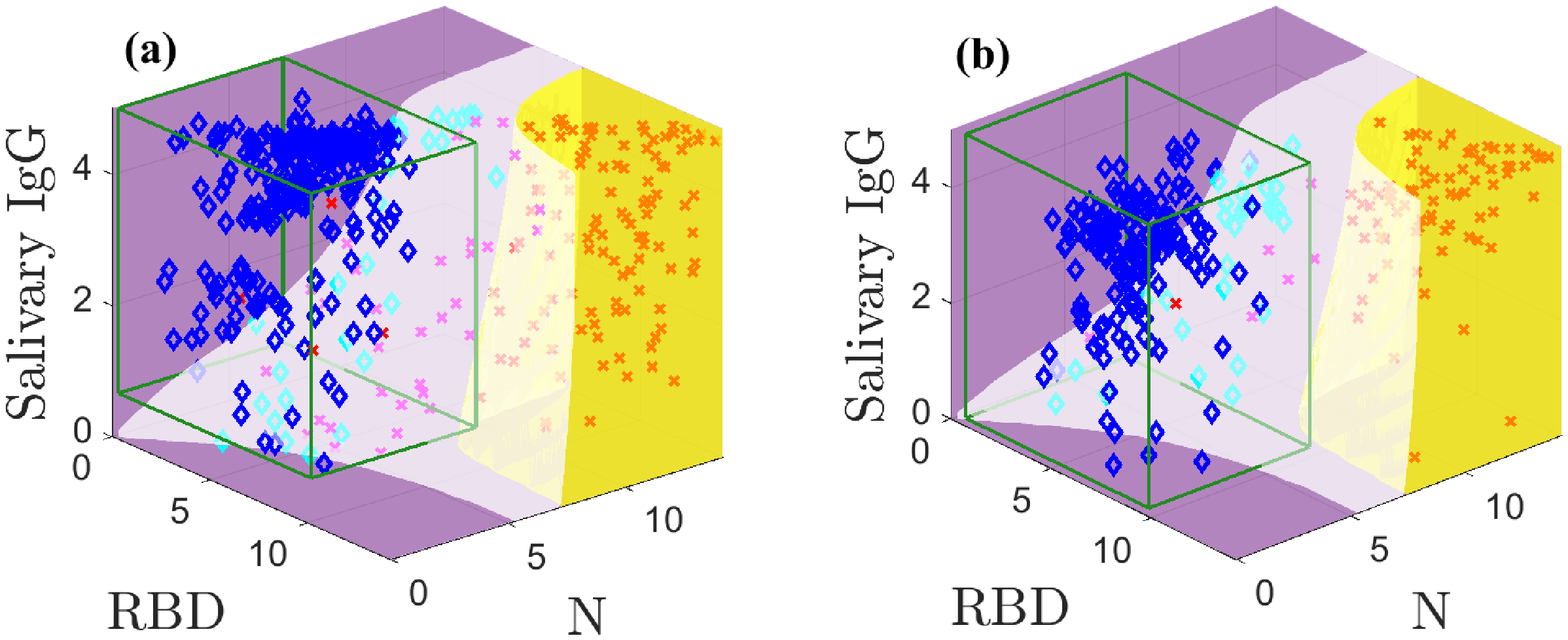}
\caption{Optimal classification domains for the training (a) and test (b) data with a holdout region (white; magenta and cyan markers are indeterminate samples). Positive samples are indicated by red \textbf{X}s and negatives with blue $\bLozenge$s. The green boxes drawn in (a) and (b) are the negative sample mean plus 3$\sigma$ confidence intervals. \href{https://livejohnshopkins-my.sharepoint.com/:f:/g/personal/rluke3_jh_edu/Emfsn6KpxtBLv9rg0VmTkFEBWn6PspYkfYTmaYyvuEJD3Q?e=eFNhOA}{Supplemental Figures 5 and 6} show animations.}
\label{fig:4}
\end{figure}

Figure \ref{fig:4}a shows the training data. Our method yields four false negatives and no false positives. In contrast, the CI yields one false positive and four false negatives. Figure \ref{fig:4}b shows the test data; the model and CI both admit one false negative but the CI allows five false positives, whereas our method correctly classifies all negative samples. 

Our classification results highlight the usefulness of working in higher dimensions while allowing flexible domains/shapes to separate data instead of CIs. Table \ref{table:1} shows our method applied to multiplex salivary SARS-CoV-2 IgG data for training and test populations \cite{pisanic2020covid,randad2021durability}. The table compares sensitivity, specificity, and classification accuracy results from our models to 3$\sigma$ CIs from the negative sample means, the original analysis \cite{pisanic2020covid}, and holdout work \cite{patrone2022holdout}. 

\begin{table}
\centering

\begin{tabular}{|lccc|}
\hline
\rowcolor[HTML]{D9D9D9} 
\multicolumn{1}{|c|}{\cellcolor[HTML]{D9D9D9}\textbf{Data and method}} & \multicolumn{1}{c|}{\cellcolor[HTML]{D9D9D9}\textbf{Positive}} & \multicolumn{1}{c|}{\cellcolor[HTML]{D9D9D9}\textbf{Negative}} & \textbf{Total} \\ \hline
\multicolumn{4}{|c|}{} \\
\multicolumn{4}{|c|}{} \\
\multicolumn{4}{|c|}{\multirow{-3}{*}{\textbf{Training samples}}}  \\ \hline
\multicolumn{1}{|l|}{} & \multicolumn{1}{c|}{\textbf{147}} & \multicolumn{1}{c|}{\textbf{283}} & \textbf{430} \\ \hline
\multicolumn{1}{|l|}{} & \multicolumn{1}{l|}{} & \multicolumn{1}{l|}{} & \multicolumn{1}{l|}{} \\
\multicolumn{1}{|l|}{\multirow{-2}{*}{\textit{\textbf{All data}}}} & \multicolumn{1}{l|}{\multirow{-2}{*}{\textbf{Sensitivity (\%)}}} & \multicolumn{1}{l|}{\multirow{-2}{*}{\textbf{Specificity (\%)}}} & \multicolumn{1}{l|}{\multirow{-2}{*}{\textbf{Accuracy (\%)}}} \\ \hline
\rowcolor[HTML]{DAE8FC} 
\multicolumn{1}{|l|}{\cellcolor[HTML]{DAE8FC}Model} & \multicolumn{1}{c|}{\cellcolor[HTML]{DAE8FC}131/147, 89.1} & \multicolumn{1}{c|}{\cellcolor[HTML]{DAE8FC}279/283, 98.6} & 410/430, 95.3 \\
\rowcolor[HTML]{FFFC9E} 
\multicolumn{1}{|l|}{\cellcolor[HTML]{FFFC9E}Confidence interval} & \multicolumn{1}{c|}{\cellcolor[HTML]{FFFC9E}121/147, 82.3} & \multicolumn{1}{c|}{\cellcolor[HTML]{FFFC9E}277/283, 97.9} & 398/430, 92.6 \\ \hline
\multicolumn{1}{|l|}{} & \multicolumn{1}{l|}{} & \multicolumn{1}{l|}{} & \multicolumn{1}{l|}{} \\
\multicolumn{1}{|l|}{\multirow{-2}{*}{\textit{\textbf{Inconclusive data excluded}}}} & \multicolumn{1}{l|}{\multirow{-2}{*}{\textbf{Sensitivity (\%)}}} & \multicolumn{1}{l|}{\multirow{-2}{*}{\textbf{Specificity (\%)}}} & \multicolumn{1}{l|}{\multirow{-2}{*}{\textbf{Accuracy (\%)}}} \\ \hline
\rowcolor[HTML]{DAE8FC} 
\multicolumn{1}{|l|}{\cellcolor[HTML]{DAE8FC}Model} & \multicolumn{1}{c|}{\cellcolor[HTML]{DAE8FC}111/115, 96.5} & \multicolumn{1}{c|}{\cellcolor[HTML]{DAE8FC}256/256, 100} & 367/371, 98.9 \\
\rowcolor[HTML]{FFFC9E} 
\multicolumn{1}{|l|}{\cellcolor[HTML]{FFFC9E}Confidence interval} & \multicolumn{1}{c|}{\cellcolor[HTML]{FFFC9E}111/115, 96.5} & \multicolumn{1}{c|}{\cellcolor[HTML]{FFFC9E}255/256, 99.6} & 366/371, 98.7 \\
\rowcolor[HTML]{C9ECC9} 
\multicolumn{1}{|l|}{\cellcolor[HTML]{C9ECC9}Pisanic et al. (2020)} & \multicolumn{1}{c|}{\cellcolor[HTML]{C9ECC9}111/115, 96.5} & \multicolumn{1}{c|}{\cellcolor[HTML]{C9ECC9}219/219, 100} & 330/334, 98.8 \\
\rowcolor[HTML]{F9D6AC} 
\multicolumn{1}{|l|}{\cellcolor[HTML]{F9D6AC}Patrone et al. (2022)} & \multicolumn{1}{c|}{\cellcolor[HTML]{F9D6AC}115/119, 96.6} & \multicolumn{1}{c|}{\cellcolor[HTML]{F9D6AC}227/227, 100} & 342/346, 98.8 \\ \hline
\multicolumn{4}{|c|}{} \\
\multicolumn{4}{|c|}{} \\
\multicolumn{4}{|c|}{\multirow{-3}{*}{\textbf{Test samples}}} \\ \hline
\multicolumn{1}{|l|}{} & \multicolumn{1}{c|}{\textbf{87}} & \multicolumn{1}{c|}{\textbf{192}} & \textbf{279} \\ \hline
\multicolumn{1}{|l|}{} & \multicolumn{1}{l|}{} & \multicolumn{1}{l|}{} & \multicolumn{1}{l|}{} \\
\multicolumn{1}{|l|}{\multirow{-2}{*}{\textit{\textbf{All data}}}} & \multicolumn{1}{l|}{\multirow{-2}{*}{\textbf{Sensitivity (\%)}}} & \multicolumn{1}{l|}{\multirow{-2}{*}{\textbf{Specificity (\%)}}} & \multicolumn{1}{l|}{\multirow{-2}{*}{\textbf{Accuracy (\%)}}} \\ \hline
\rowcolor[HTML]{DAE8FC} 
\multicolumn{1}{|l|}{\cellcolor[HTML]{DAE8FC}Model} & \multicolumn{1}{c|}{\cellcolor[HTML]{DAE8FC}83/87, 95.4} & \multicolumn{1}{c|}{\cellcolor[HTML]{DAE8FC}189/192, 98.4} & 272/279, 97.5 \\
\rowcolor[HTML]{FFFC9E} 
\multicolumn{1}{|l|}{\cellcolor[HTML]{FFFC9E}Confidence interval} & \multicolumn{1}{c|}{\cellcolor[HTML]{FFFC9E}81/87, 93.1} & \multicolumn{1}{c|}{\cellcolor[HTML]{FFFC9E}188/192, 97.9} & 269/279, 96.4 \\ \hline
\multicolumn{1}{|l|}{} & \multicolumn{1}{l|}{} & \multicolumn{1}{l|}{} & \multicolumn{1}{l|}{} \\
\multicolumn{1}{|l|}{\multirow{-2}{*}{\textit{\textbf{Inconclusive data excluded}}}} & \multicolumn{1}{l|}{\multirow{-2}{*}{\textbf{Sensitivity (\%)}}} & \multicolumn{1}{l|}{\multirow{-2}{*}{\textbf{Specificity (\%)}}} & \multicolumn{1}{l|}{\multirow{-2}{*}{\textbf{Accuracy (\%)}}} \\ \hline
\rowcolor[HTML]{DAE8FC} 
\multicolumn{1}{|l|}{\cellcolor[HTML]{DAE8FC}Model} & \multicolumn{1}{c|}{\cellcolor[HTML]{DAE8FC}80/81, 98.8} & \multicolumn{1}{c|}{\cellcolor[HTML]{DAE8FC}163/163, 100} & 243/244, 99.6 \\
\rowcolor[HTML]{FFFC9E} 
\multicolumn{1}{|l|}{\cellcolor[HTML]{FFFC9E}Confidence interval} & \multicolumn{1}{c|}{\cellcolor[HTML]{FFFC9E}80/81, 98.8} & \multicolumn{1}{c|}{\cellcolor[HTML]{FFFC9E}158/163, 96.9} & 238/244, 97.5 \\
\rowcolor[HTML]{C9ECC9} 
\multicolumn{1}{|l|}{\cellcolor[HTML]{C9ECC9}Pisanic et al. (2020)} & \multicolumn{1}{c|}{\cellcolor[HTML]{C9ECC9}81/81, 100} & \multicolumn{1}{c|}{\cellcolor[HTML]{C9ECC9}125/126, 99.2} & 206/207, 99.5 \\
\rowcolor[HTML]{F9D6AC} 
\multicolumn{1}{|l|}{\cellcolor[HTML]{F9D6AC}Patrone et al. (2022)} & \multicolumn{1}{c|}{\cellcolor[HTML]{F9D6AC}81/82, 98.8} & \multicolumn{1}{c|}{\cellcolor[HTML]{F9D6AC}157/158, 99.4} & 238/240, 99.2 \\ \hline
\end{tabular}
\caption{
Summary information about the SARS-CoV-2 datasets with sensitivities, specificities, and classification accuracies for training and test data with and without allowing an indeterminate class.  Model and 3$\sigma$ (relative to negative sample mean) confidence interval results are shown for all samples; the 
original analysis \cite{pisanic2020covid} was conducted on all seven antibody targets and the ELISA-based total IgG without indeterminate samples.}
\label{table:1}
\end{table}

Our model-based classification accuracy (prevalence-weighted sum of sensitivity and specificity) always improves upon that given by CIs. We achieve classification error rates under 1.1 \% in two cases, which are not attained by the CIs for any subpopulation. On average, we reduce the classification error by 41.6 \%, with an 84.0 \% reduction for the test data with an indeterminate class. Overall errors are larger without an indeterminate class; this shows the challenges of working with overlapping positive and negative data.  

	Table \ref{table:1} also compares our model-based accuracy to the original analysis \cite{pisanic2020covid}, which created their own indeterminate class. Using a local accuracy threshold of 99 \%, we reduce the number of indeterminate samples by 28.7 \% for the training data and by 51.6 \% for the test data. The original analysis \cite{pisanic2020covid} used the ELISA-based salivary total IgG and all seven antibody targets, although the latter were summed to form a single number. Thus, while the original work \cite{pisanic2020covid} required eight dimensions (i.e., measurements), the analysis was projected onto two. In contrast, we only use two antibody targets and the ELISA-based salivary total IgG, treating these as a measurement triple. We improve classification accuracy from 98.8 \% to 98.9 \% for the training data and from 99.5 \% to 99.6 \% for the test data. While the improvements are small, we achieve them \textit{while using 13.8 \% more of the available data}. This example highlights the ability of modeling to simultaneously minimize classification errors, use fewer antibody targets, and hold out fewer samples.
	
	In Table \ref{table:1} we also compare sensitivity and specificity results from our models against corresponding results from 3$\sigma$ CIs. Our model sensitivities and specificities match or best those given by CIs. We increase sensitivities from 92.7 \% to 95.0 \% and specificities from 98.1 \% to 99.3 \% on average, \textit{even including the data in our indeterminate region}. Excluding indeterminate data, we achieve 100 \% specificity for both the training and test populations, which is not realized by CIs. The number of false positives and false negatives increases when indeterminate samples are considered. However, we still have fewer false classifications as compared to CIs.
		
Table \ref{table:1} includes sensitivities and specificities from the original analysis \cite{pisanic2020covid}. Our approach matches training data sensitivities and specificities of 96.5 \% and 100 \%. Test data sensitivity is lowered from 100 \% to 98.8 \% but specificity is increased from 99.2 \% to 100 \%. This discrepancy arises from the different objectives in our work versus the original analysis \cite{pisanic2020covid}; our optimal classification maximizes the prevalence-weighted sensitivity and specificity, whereas the prior method maximized specificity while maintaining an acceptable sensitivity. As such, our results do not outperform the original analysis for each case, but we improve overall classification accuracy. Further, our method yields acceptable sensitivities and specificities using the indeterminate data in our holdout region; the original analysis excluded those samples.

	We also report sensitivity, specificity, and classification accuracy results from holdout work \cite{patrone2022holdout} in Table \ref{table:1} for comparison. We reduce the number of holdout samples by 14.6 \% relative to their method. In doing so, we use 25 more training samples while holding fixed the number of misclassified points at four. Further, we use four additional test samples and have one fewer misclassifications, thereby reducing test classification error by 50 \%. This improvement is surprising given that their original work already significantly decreased the number of holdouts. Like the original analysis of the data \cite{pisanic2020covid}, that study \cite{patrone2022holdout} summed seven of the eight available antibody measurements to yield a single number; our results suggest that considering measurements separately improves performance.

\section{Discussion}
 
High dimensional mathematical modeling is a powerful tool for classification. Adding dimensions can improve data separation and allow models to better leverage the underlying structure, thereby increasing classification accuracy. We illustrate high dimensional modeling with a binary classification of 3D SARS-CoV-2 antibody response measurements. Our 3D modeling yields significant improvements over CI methods, even when indeterminate data are considered. For the examples considered herein, we decrease average classification errors by 41.6 \%. A combined specificity + sensitivity of at least 150 \% is desirable \cite{power2013principles}; our average (194 $\pm$ 4.79) \% is close to the perfect 200 \% and higher than that of CIs (191 $\pm$ 7.40) \%. 

Our work is limited by the inherent subjectivity of selecting mathematical models. In high dimensions, new structures of the data become apparent, which can necessitate more nuanced modeling. However, this does not suggest how to determine when a model is optimal. One approach is to construct a family of models and identify the one with the smallest error rate \cite{patrone2021classification}. Encouragingly, the subjectivity issue may be lessened by adding more data points \cite{schwartz1967estimation}. Our models are not designed to account for waning antibody levels, which can drop below detection thresholds when measured several months after infection \cite{xia2021longitudinal}. Further, our models do not differentiate between demographic factors like age or biological sex, although data could be stratified based on these factors. For example, children have better COVID-19 outcomes than adults \cite{yuki2020covid}, suggesting the usefulness of modeling their antibody test results as a distinct group. 

A significant benefit of our modeling approach is its adaptability to any number of measurement targets and classes. While our work is demonstrated using N, RBD, and ELISA-based salivary total IgG, modeling four targets and beyond is possible. A more challenging extension is to identify which antibody targets minimize classification errors; we anticipate difficulties in visualization and comparison between dimensions. To this end, models could be constructed for all possible antigen and ELISA combinations to determine the result yielding the highest classification accuracy. More generally, our method does not address the problem of antigen down-selection, which is important for assay design. Given the competing needs to understand correlates of protection and estimate prevalence, it is unclear how to define an objective function that yields an optimal choice of antigens. We leave this for future work.

Finally, our classification scheme minimizes average rates of false negatives and false positives, which relate to sensitivity and specificity. The problem could be formulated as a constrained optimization to meet desired sensitivity and specificity targets to create a ``rule-out'' or ``rule-in'' test \cite{florkowski2008sensitivity}. 
	
	In conclusion, the usefulness of our procedure is due to the inherent separation of data in additional dimensions and the ability of models to fit the structure of the data, which molds the classification domains to the negative and positive populations. Our high dimensional models achieve superior classification accuracy and have the potential to replace traditional methods. 
	
	\section{Acknowledgements}
	
The authors wish to thank Dr. Daniel Anderson and Erica Ramsos for useful feedback during preparation of the manuscript.

This work is a contribution of the National Institutes of Standards and Technology and is therefore not subject to copyright in the United States.

Use of data provided in this manuscript has been approved by: (1) the NIST Research Protections Office (IRB no. ITL-2020-0257); and (2) the Johns Hopkins School of Medicine Institutional Review Board (IRB no. IRB00247886).

\section{Declaration of Competing Interest}

RL, AK, NP, CH, and PP have no conflicts of interest to declare.

YM has received consulting fees from Abbott Diagnostics; served leadership or fiduciary roles in the Infectious Diseases Institute, Infectious Diseases Society of America, and WHO TDR; and her institution received reagents from Hologic, Chembio, Roche, Cepheid, and Becton-Dickinson for studies.

DT has received funds for Merck DSMB board membership, consulting fees and stock or stock option for Excision Bio, fees for expert testimony, honoraria for CME programs only, and royalties from UpToDate.

\section{Funding} 

Funding for Johns Hopkins University authors was provided by the Johns Hopkins COVID-19 Research and Response Program, the FIA Foundation, a gift from the GRACE Communications Foundation (C.D.H., N.P.), National Cancer Institute (NCI) SeroNet grant U01CA260469 (C.D.H.), National Institute of Allergy and Infectious Diseases (NIAID) grant R21AI139784 (C.D.H. and N.P), National Institute of Environmental Health Sciences (NIEHS) grant R01ES026973 (C.D.H., N.P.), NIAID grant R01AI130066 and NIH grant U24OD023382 (C.D.H), NIAID grant 3R01AI148049 (D.L.T.), the Johns Hopkins University School of Medicine COVID-19 Research Fund, the Sherrilyn and Ken Fisher Center for Environmental Infectious Diseases Discovery Program, and NIH grants U54EB007958-12, U5411090366 (Y.C.M.). The aforementioned funders had no role in study design, data analysis, decision to publish, or preparation of the manuscript. R.L. was also funded through the NIST PREP grant 70NANB18H162.

\setcounter{theorem}{0}
\setcounter{equation}{0}

\appendix

\renewcommand{\theequation}{A\arabic{equation}}

\section{Details of Models}
\label{sec:app_a}

Our mathematical models are probability distributions that describe the chances of observing a positive or negative sample in 3D measurement space. To create a probability distribution, we select a parameterized model that qualitatively describes the shape of the population. Examples of distributions commonly used to model biological phenomena include the normal, uniform, beta, and exponential. 

To motivate a specific choice, first note that much of the negative data have N and RBD values that are roughly proportional. Mathematically, this means that the data lies along the diagonal line $y = x$, suggesting the change of variables 
\begin{equation}
u = \frac{x + y}{\sqrt{2}}, \quad w = \frac{x  - y}{\sqrt{2}}, \quad v = z.
\end{equation}
Additionally, the negative data fans out away from the origin along the diagonal (see Figure \ref{fig:1}a). This suggests that the variance of the difference $w$ between N and RBD increases with their sum, i.e., the variable $u$. We empirically choose the variance $\sigma_w$ of this difference to be
\begin{equation}
\sigma_w = \alpha \exp \left[ \frac{u - \mu_u}{\beta} \right],
\label{eq:a2}
\end{equation}
for constants $\alpha$ and $\beta$, where $\mu_u$ is a characteristic total SARS-CoV-2 antibody level, which is yet to be determined.

A 3D distribution in variables $u, w$, and $v$ was created for the negative population. We select a hybrid triple normal distribution having the form
\begin{equation}
N(\bm{r}) = \frac{1}{(2 \pi)^{3/2} \sigma_u \sigma_w \sigma_v} \exp \left\{ - \frac{1}{2} \left[ \left( \frac{u - \mu_u}{\sigma_u} \right)^2 + \left( \frac{w - \mu_w}{\sigma_w} \right)^2 + \left( \frac{v - \mu_v}{\sigma_v} \right)^2 \right] \right\},
\end{equation}
where $\sigma_u$ and $\sigma_v$ are constants and $\sigma_w$ is defined by Eq. \ref{eq:a2}. We use maximum likelihood estimation (MLE) to identify the model parameters $\mu_u, \mu_w, \mu_v, \sigma_u, \sigma_v, \alpha$, and $\beta$ that maximize the probability of observing the negative training data \cite{williams2006gaussian}.

Similarly, we introduce changes of variables and model the positive population. The structure of the positive data motivates a change to a spherical coordinate system:
\begin{equation}
\zeta = \sqrt{x^2 + y^2 + z^2}, \quad \omega = \arctan \left(  \frac{\sqrt{x^2 + y^2}}{z}\right), \quad \phi = \arctan \left( \frac{y}{x}\right), x > 0.
\end{equation}
This reflects the fact that the data moves out radially in 3D from the origin. We use a weighted hybrid triple normal distribution:
\begin{equation}
P(\bm{r}) = \frac{1}{4 \pi^{3/2} \sigma_{\zeta} \sigma_{\omega} \sigma_{\phi}} \exp \left\{ - \frac{1}{2} \left[ \left( \frac{\zeta - \mu_{\zeta}}{\sigma_{\zeta}} \right)^2 + \left( \frac{\omega - \mu_{\omega}}{\sigma_{\omega}} \right)^2 + \frac{1}{2} \left( \frac{\phi - \mu_{\phi}}{\sigma_{\phi}} \right)^2 \right] \right\},
\end{equation}
where $\sigma_{\zeta},\sigma_{\omega}$ and $\sigma_{\phi}$ are constants. MLE is used to determine the optimal model parameters using the positive training data.

	Using either a known prevalence $q$ or estimated prevalence $\hat{q}$ (see Section \ref{sec:app_b}), we then classify a measurement $\bm{r}$ as positive if
\begin{equation}
(1-q)N(\bm{r})<qP(\bm{r})					
\end{equation}
and negative if
\begin{equation}
qP(\bm{r})<(1-q)N(\bm{r}). 					
\end{equation}

\section{Prevalence Estimation}
\label{sec:app_b}

We can estimate the disease prevalence if it is unknown following \cite{patrone2021classification}. The strategy  requires dividing the measurement space into two arbitrary regions. Many clustering algorithms to select these regions exist \cite{abbas2008comparisons,jain2010data}; our implementation chooses these regions based on a $k$-means clustering of the data. For a binary classification, we use $k=2$ clusters and assign each point in our 3D measurement space to the cluster with the closest mean. We select a region $D$ on one side of the boundary separating the clusters and count the total number of points irrespective of their (unknown) class; call this number $Q_D$. We compute the integrals
\begin{equation}
P_D = \int_D P(\bm{r}) d \bm{r}, \quad N_D = \int_D N(\bm{r}) d \bm{r},
\end{equation}
which are the probabilities that a known positive or known negative sample falls in the domain $D$. Then our prevalence estimate is computed as
\begin{equation}
\hat{q} = \frac{Q_D - N_D}{P_D - N_D}.
\end{equation}
This estimate $\hat{q}$ can be used in place of the true prevalence when $q$ is unknown. Our estimate is unbiased. Moreover, as the number of samples increases, $\hat{q}$ will converge to $q$.

\section{Local Accuracy}
\label{sec:app_c}

The local accuracy, $Z$, of a test sample with measurement $\bm{r}$ is given by 
\begin{equation}
Z_P (\bm{r})=\frac{qP(\bm{r})}{qP(\bm{r})+(1-q)N(\bm{r})}	
\label{eq:app10}
\end{equation}
if the sample falls in the optimal positive domain, and by
\begin{equation}
Z_N (\bm{r})=\frac{(1-q)N(\bm{r})}{qP(\bm{r})+(1-q)N(\bm{r})}	
\label{eq:app11}
\end{equation}
if the sample falls in the optimal negative domain. The denominators in Eqs. \ref{eq:app10} and \ref{eq:app11} give the probability that a test sample has a measurement value $\bm{r}$. See \cite{williams2006gaussian} for details.

\bibliographystyle{unsrt}
\bibliography{Multiclass_bib}

\end{document}